# Morphology of Nanoclusters and Nanopillars Formed in Nonequilibrium Surface Growth for Catalysis Applications

Vyacheslav Gorshkov,[a,b] Oleksandr Zavalov,[a]
Plamen B. Atanassov[c] and Vladimir Privman[d,*]

[a] Institute of Physics, National Academy of Sciences, 46 Nauky Avenue, Kiev 03028, Ukraine
[b] National Technical University of Ukraine — KPI, 37 Peremogy Avenue, Building 7, Kiev 03056, Ukraine
[c] Center for Emerging Energy Technologies, Department of Chemical and Nuclear Engineering, University of New Mexico, Albuquerque, NM 87131, USA
[d] Center for Advanced Materials Processing, Department of Physics, Clarkson University, Potsdam, NY 13699, USA

## Abstract

We consider growth of nanoclusters and nanopillars in a model of surface deposition and restructuring yielding morphologies of interest in designing catalysis applications. Kinetic Monte Carlo numerical modeling yields examples of the emergence of FCC-symmetry surface features, allowing evaluation of the fraction of the resulting active sites with desirable properties, such as (111)-like coordination, as well as suggesting the optimal growth regimes.

---

[*] Corresponding author: e-mail privman@clarkson.edu; phone +1-315-268-3891

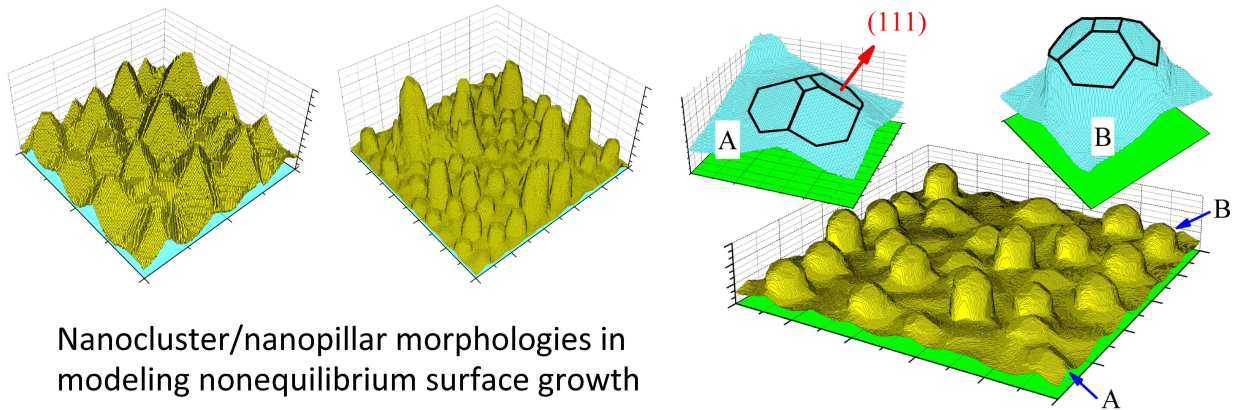

Nanocluster/nanopillar morphologies in modeling nonequilibrium surface growth

Web-link to future updates of this article:  www.clarkson.edu/Privman/230.pdf

**I. Introduction**

Nanosize morphology, especially its emergence in surface growth by processes of attachment and restructuring of deposits formed by atoms, ions or molecules, is an active field of research[1-7] driven by numerous applications. For growth on surfaces utilized as supports in catalysis, emergence of the appropriate-symmetry crystalline faces of enhanced activity, as part of the exposed on-surface deposit, for instance, (111) for Pt, is of importance. Recently, it has been experimentally demonstrated[8-10] that nanoclusters and nanopillars can be formed in surface growth, including structures with a substantial fraction of (111) faces. These findings pose new theoretical challenges. Specifically, practical modeling approaches are needed to help answer question such as which substrates are the best for growing such morphologies, as well as what is the optimal amount of matter to be deposited to have the maximal (111) or other preferred orientation face area. Other issues to explore include the dependence of the growth process on the physical conditions: temperature, flux of matter, etc.

Thus, we are interested in studying how can surface structures be grown with well-defined, preferably uniform morphology of the emerging features: nanoclusters or larger nanopillars. This calls for a general modeling approach focusing on the shape-selection resulting from the competition of several dynamical processes: transport of matter, on-surface restructuring, and detachment/reattachment. We do not attempt to consider the properties of these structures relevant for their actual use once synthesized, for instance, the chemistry and physics of their catalytic activity. Therefore, a generalized approach focusing on the kinetics of the constituent building blocks: atoms, ions or molecules, to be termed "atoms" for brevity, is feasible. In fact, such a model has recently been developed[11] for the unsupported (off-surface) growth of nanoparticles of well-defined shapes. In the present work we adapt this approach to surface-feature formation.

One important recent theoretical observation[11-13] has been that "persistency" can be a driving mechanism in the emergence of well-defined shapes in nonequilibrium growth of nanoparticles and nanostructures. Initially, the property termed imperfect-oriented attachment[14-17] has been identified as persistency in successive nanocrystal binding events leading to the formation of uniform short chains of aggregated nanoparticles. Persistency can also mediate growth of other shapes[11-13,17] from atoms, for a certain range of the resulting particle and feature sizes. Indeed, nanosize particles and structures, for many growth conditions are simply not sufficiently large (do not contain enough constituent atoms) to develop large internal defects and unstable surface features that result in the formation of whiskers and/or the "dendritic instabilities" of growing side branches, then branches-on-branches, etc. — processes which can distort a uniform shape with approximately crystalline faces to cause it to evolve into a random/fractal or snowflake like morphology.[18,19]

We have to consider several processes and their competition, which together result in the deposit morphology and feature-shape selection. We will assume diffusional transport followed by attachments of the atoms (ions, molecules) to the growing surface deposit. Furthermore, these atoms can detach and reattach. They can also move and roll on the surface, according to thermal-like rules which will be detailed later. Indeed, diffusional transport without such restructuring would yield a fractal structure.[18,19] Moreover, particle and surface-structure synthesis is typically carried out in the nonequilibrium regime, the definition of which has recently been explored in isolated nanoparticle growth.[11] When the outer surface layer relaxation processes occur on time scales, $\tau_r$, much shorter that the time scales, $\tau_d$, of the formation of additional layers due to diffusional transport of matter to the surface, then steady-state growth is expected. In this regime, $\tau_r \ll \tau_d$, isolated particles assume Wulff shapes.[20-23] For surface growth this regime is not particularly interesting in the present context. Recall that we are not considering the scaling-limit, unlimited amount of matter type growth, but only an overgrowth of the initial substrate with a finite number of surface layers of nanosize thickness. In the opposite limit, $\tau_r \gg \tau_d$, growth becomes irregular, and no definite nanoscale particle[11] or surface-feature shapes are expected.

Here we study the practically important nonequilibrium regime, $\tau_r \sim \tau_d$, for which emergence of well-defined nanocrystal shapes has recently been demonstrated by the kinetic Monte Carlo (MC) type approach considered here, for isolated particle growth,[11] for the main crystalline symmetries. This approach, introduced in Section II, requires substantial numerical resources, but offers the flexibility of allowing to explicitly control the kinetics of atoms hopping on the surface and detaching/reattaching, according to thermal-type, (free-)energy-barrier rules. The diffusional transport occurs in the three-dimensional (3D) space. However, the atom attachment is only allowed "registered" with the underlying lattice of the initial substrate. The latter rule prevents the growing structures from developing "macroscopic" (structure-wide) defects, which has been a property identified[11] as important for well-defined particle shape selection in isolated particle growth, with shapes defined by faces of the crystalline symmetry of the substrate, but with proportions different from those in the equilibrium Wulff growth. For example, for the simple-cubic (SC) lattice symmetry, a cubic shape nanoparticle can only be obtained in the nonequilibrium regime, whereas the Wulff shapes are typically rhombitruncated cuboctahedra. The type of results of interest here are illustrated in Figure 1, which shows a "time series" of numerically grown nanocluster morphologies: panels (a-d), as well as a snapshot of a single nanopillar: panel (e), obtained in the nonequilibrium regime for the FCC lattice structure. The parameters of the growth model are defined later, in Section II.

The rest of this article is organized as follows. In Section II, we present details of the numerical approach, accompanied by a discussion of the physical interpretation of the assumptions and dynamical rules used. Section III is devoted to discussion of the results and to concluding remarks.

## II. The Model and Its Numerical Implementation

The selection of the substrate for deposition of catalytically active structures, is an important problem on its own. Both the crystallographic orientation of the substrate and its detailed structure if not flat (for seeding/templating), can affect the morphology of the resulting deposit. The present study is aimed at an illustration of the capability of the developed kinetic MC approach to reproduce well-defined nanocluster/nanopillar surface morphologies. Since numerical simulations are resource-demanding, in this work we will consider a specific, relatively generic example with the FCC symmetry, though most of the discussion is more general. For the substrate, we select an ideal, flat (100) lattice plane. This choice is further discussed later in this section.

The pointlike building-block "atoms" undergo free, continuous-space (off-lattice) Brownian motion. They can be captured into vacant lattice sites adjacent to the growing structure, according to the following rule. Each vacant lattice site which is a nearest-neighbor of at least one occupied site is surrounded by a conveniently defined "box" (we used the Wigner-Seitz unit-lattice cell). If an atom moves to a location within such a box, it is captured and positioned exactly at the lattice location in it center. The on-surface restructuring rules, detailed below, are also such that the precise "registration" with the lattice is maintained. For simplicity, the original substrate atoms are kept fixed. All the other atoms can not only move on the surface but also detach. As mentioned in the Introduction, the precise lattice-"registration" property is crucial[11] for the emergence of the morphologies of interest, because while allowing the formation of voids in the growing deposit, we thereby prevent the formation of defects of the type that can have a "macroscopic" effect in that they can dominate the dynamics of the particle/feature growth as a whole, for instance, by preferentially driving the growth of certain faces or sustaining unequal-proportion shapes. Nanocrystal and surface morphologies of interest here, in most cases are obtained in the regime where such defects are dynamically avoided/dissolved, which is mimicked by our "exact registration" rule.

Let us now briefly outline our numerical implementation of the off-lattice diffusion, the details of which are given elsewhere.[11] Each diffusing atom hops a distance $\ell$ per each time step, in a random direction. Typically, $\ell$ is set to the *cubic* lattice spacing (of FCC), and hopping attempts into any aforedefined cell which already contains an occupied lattice site at its center, are failed. We will assume units such that both the time step of each MC "sweep" through the system and the distance $\ell$ are set to 1, Then in these dimensionless units the diffusion constant[11] is $D = 1/6$.

The actual dynamics is carried out in a box-shaped region typically taken of dimensions $X \times Y \times Z$ up to $500 \times 500 \times 500$ (each in units of $\ell$). The initial substrate is at the bottom of the box, at $z = 0$. Periodic boundary conditions are used in both horizontal directions, $0 \leq x < X$ and $0 \leq y < Y$. In the course of the simulation, the total count of particles in the topmost boundary layer of thickness 4, located at $Z - 4 \leq z < Z$, is monitored and, as it changes, particles are replenished (at random locations in that layer) or removed to maintain their total number, *N*, constant, with the density of

$$n = N/4XY,  \qquad (1)$$

in units of $\ell^{-3}$. The rest of the box, for $0 < z < Z - 4$, is initially empty. Simulations were carried out for several horizontal box sizes, to check that there was no size-dependence of the results.

Unlike the horizontal box dimensions, the vertical size, *Z*, can only be adjusted with care. Indeed, after a short, but nonnegligible, transient time which is typically a sizable fraction of the time it takes the first on-surface clusters to form, see Figure 1, panel (a), the density distribution in the box reaches an *approximately* linear one, $\approx nz/Z$, and the flux of matter to the surface, $\Phi$, assumes an approximately steady-state value of $\Phi \approx Dn/Z$, which is *Z*-dependent. All our simulations corresponded to the growing structures remaining at least at the distance of ~ 200 units away from the topmost boundary layer. We kept *Z* = 500, and no attempts were made to otherwise maintain the flux stationary, or have a more "realistic" time-dependence as the structures grew. Thus, the diffusional supply of matter (the flux of atoms to the growing surface), while somewhat geometry- and time-dependent, is, at least initially, for approximately steady state conditions that are rapidly achieved, proportional to the product *Dn*. It is therefore one of the physical parameters of the growth process that can be modified, e.g., by adjusting *n*, or even made manifestly time-dependent, by varying *n*(*t*), to control the resulting deposit morphology. In our simulations, however, *n* was kept constant, and the process was simply stopped after a selected time, *t*. The time of the growth, *t*, is, in fact, another physical parameter that allows control of the resulting structure.

The deposited atoms in the growing structure are not fixed. They can hop to nearby vacant lattice sites without losing contact with the main structure, or actually detach, thus rejoining the "free" atom population. Here the set of possible displacement vectors, $\vec{e}_i$ (if the target site was vacant), included only those pointing to the nearest neighbors. However, inclusion of next-nearest-neighbor displacements was considered in modeling isolated nanoparticle growth by this approach, and is known to have an effect on the nanocluster shape proportions.[11] The specific dynamical rules here follow those in the earlier work,[11] but they should not be taken too literally, because they only mimic thermal-type transitions and are not corresponding to any actual physical interactions, for instance those of Pt atoms, nor to any realistic kinetics. The reason for such an approach has been that more realistic modeling would require prohibitive numerical resources and thus make it impractical to study large enough systems to observe the features of interest in surface structure morphology formation.

We now turn to the description of on-surface motion and possible detachment of atoms. Each atom with at least one vacant neighbor site (means, capable of moving) will have a coordination number $m_0 = 1, \ldots, 11$ (for nearest-neighbor FCC). In each MC sweep through the system (unit time step), in addition to moving each free atom, we also attempt to move each attached (surface) atom that has vacant neighbor(s), except those atoms which are in the original flat substrate and are immobile. We assume that the probability for a surface atom to actually move during a time step is given by $p^{m_0}$, i.e., that there is a certain (free-)energy (per $kT$) barrier, $m_0 \delta > 0$, to overcome, so that $p \sim e^{-\delta} < 1$. However, the probability for the atom, if it moves, to hop to any of its $12 - m_0$ vacant neighbor sites will be assumed not uniform but proportional to $e^{m_i |\varepsilon|/kT}$, where $\varepsilon < 0$ is a certain free-energy at the target site, the final-state coordination of which, if selected and occupied, will be $m_i = 0, \ldots, 11$. Typically, our simulations have involved up to $3 \times 10^7$ unit-time MC sweeps, corresponding to the MC dimensionless "time" of growth, $t$, with the total number of deposited atoms up to $1.5 \times 10^7$.

On-surface atom motion and detachment generally involve at least two physical parameters: the surface diffusion constant, $D_s$, and the temperature, $T$. As typical for such "cartoon" models of particle deposition kinetics, our transition rules are not directly related to realistic atom-atom and atom-environment interactions or entropic effects, and, given that we are studying a nonequilibrium regime, no attempt has been made to ensure thermalization (to satisfy detailed balance, for instance). However, loosely we expect that $D_s$ is related to $p$, is temperature-dependent, and reflects the surface-binding energy barriers. The other parameter to vary in order to mimic the effects of changing the temperature, is

$$\alpha = |\varepsilon|/kT ,\qquad(2)$$

which involves (free-)energy scales, $\varepsilon$, more related to the entropic properties. These expectations are primarily based on empirical observations. Indeed, we found that the parameter $p$ is best kept approximately in the range 0.6-0.7, which seems to correspond to the nonequilibrium growth[11] with $\tau_r \approx \tau_d$, as defined in the Introduction. The parameter $\alpha$ should be in the range 1-3 for interesting shapes to emerge.

There are obviously other "microscopic" parameters in the problem that can be adjusted, such as, for instance, the attachment probability of the arriving atoms, which could be made less than 1, etc. However, there are also "macroscopic" parameters, such as the geometry of the system and the lattice symmetry-related properties, that can also be modified. One important choice is that of the initial substrate for deposition. Growth of isolated nanoparticles, considered in an earlier study,[11] yields useful insights into the problem of selecting a substrate for surface-feature formation. Specifically, in the nonequilibrium regime, nanosize shapes can be dominated, for a range of growth times and particle sizes, by densely packed faces of symmetries similar to those encountered in the Wulff construction, but with different proportions. For FCC, Figure 2 shows the Wulff form, involving the (100) and (111) type faces. For nonequilibrium growth, two shapes are shown. One still has the (100) and (111) faces, but in the other, formed under faster-growth conditions, the (111) faces "win" and dominate the shape. Generally, this suggests that (100) and (111) are naturally complementary lattice faces in nonequilibrium FCC-symmetry growth, and therefore (100)-consistent substrates are a good choice for growing (111), if a (111)-consistent initial surface is not an option. Similar examples for some other lattice symmetries are available.[11] In addition to these empirical/theoretical considerations, we note that the natural emergence of octahedral shapes made of (111)- and (100)-type faces, for on-surface Pt nanoclusters, is a well established property in experiment.[24]

**III. Results and Discussion**

A key finding of the present work has been the mere observation that relatively simple and controllable kinetic models can yield growth modes with the formation of well-defined surface structures resembling nanoclusters and nanopillars similar to those observed in recent experiments. However, a more detailed analysis of the growth process is possible and addressed in this section. Specifically, we will consider the structural properties of the formed deposit: the nanocluster/nanopillar formation and sizes. We will also explore the fraction of the grown structure which has surface atoms in the desirable, here (111) type configuration. Matters related to possible "optimization" of the growth process will be discussed.

Figure 3 illustrates a typical structure with nanocluster "pyramids" grown initially, which for larger times develop into a collection of nanopillars of a broader size distribution; this is also seen in Figure 1, panels (b-c), for another set of parameters. The emergence of such morphologies requires a proper selection of the growth parameter values, because general parameter choices typically just lead to a random surface growth. In order to initiate cluster formation, islands must first form on the flat substrate, e.g., Figure 1, panel (a), and then act as seeds for cluster growth rather than merge with each other. The kinetics of the initial, few-layer cluster size distribution, is controlled by the on-surface restructuring process rates, set by parameters such as the surface diffusion constant, $D_s$, mentioned earlier, but also by the incoming flux, $\Phi \approx Dn/Z$, discussed in Section II. The latter is the easiest to control experimentally, for instance by varying n, and we found empirically that, $d \sim n^{-1/3}$, up to those times for which pyramidal shapes are obtained. We can devise arguments for this relation, but they are too speculative to detail here. Instead, we offer numerical evidence, see Figure 4, as discussed in the next two paragraphs. Note that the proportionality coefficient in the relation $d \sim n^{-1/3}$ is not only dimensional but, even in dimensionless units, is well over 1. Since $n$ always enters via the (dimensional) flux, $\Phi \approx Dn/Z$, it follows this coefficient is $Z$- and (weakly) time-dependent.

The effective cross-sectional dimension of the growing clusters, $d(t)$, was actually numerically estimated by calculating the height-height correlation function,

$$G(\Delta x, \Delta y, t) = \frac{\langle \Delta z(x, y, t) \Delta z(x + \Delta x, y + \Delta y, t) \rangle}{\langle [\Delta z(x, y, t)]^2 \rangle}, \qquad (3)$$

where $\Delta z(x, y, t) = z(x, y, t) - \langle z(x, y, t) \rangle$, and the averages, $\langle \cdots \rangle$, are over all the $(x, y)$ substrate coordinates. This correlation function is, as usual, oscillatory in the distance from the origin of the $(\Delta x, \Delta y)$ horizontal displacement plane, and, for simplicity, we define $d(t)$ as the location of its first zero along the $\Delta x$-direction: $G(d(t), 0, t) = 0$. (Here the coordinates are considered continuous: the function was linearly extrapolated to noninteger $\Delta x$ values.)

The dynamics of the surface morphology evolution proceeds as follow. Initially, nanoclusters emerge, forming mostly independently, with their structure developing similarly to that of clusters in Figure 2: While significant randomness and fluctuations are present, generally the nanoclusters develop into (slightly truncated) pyramids shaped as halves of clusters grown as isolated entities, illustrated in the lower panels of Figure 2. The transverse dimension of the resulting structures, measured by $d(t)$, evolves as shown in Figure 4. There is a certain time interval during which $d(t) \approx d_{st}$ is approximately constant (a "plateau" region), and the nanoclusters evolve by developing characteristic, approximately uniform pyramidal shapes. At later times, the clusters begin to compete with one another, by coarsening partly at the expense of each other and by larger clusters screening the growth of the small ones. The resulting morphology is then that of nanopillars, but their size distribution is not narrow; they show significant variation in both height and girth.

In isolated cluster growth,[11] the "persistence" in the cluster morphology evolution, mentioned in the Introduction, has resulted in a regime of times during which relatively well-defined shapes, such as those in Figure 2, were formed. However, for larger times particle shapes will eventually destabilize and become random or dendrite-like. Here, for on-surface growth, the shorter-time regime of the isolated cluster formation is also present. Our numerical results, such as those shown in Figures 1 and 3, suggest that yet another regime, that of competing nanopillars, is found. Eventually, for large times, for diffusional transport of matter to the surface we expect that this growth morphology will also destabilize and the structure will ultimately become more random (e.g., fractal). Most of our simulations did not go to large enough times to see this regime, the onset of which can be seen Figure 1, panel (d). Furthermore, the selected cluster shown in panel (e) — one of the larger nanopillars obtained in the regime illustrated in panel (c) — demonstrates the onset of the screening as it even effects this nanopillar's own lower section, which is narrower than its top section. Note that most nanopillars in panel (c) are still in the regime of having broader base sections than their top sections. For larger times the broader top section will ultimately begin to destabilize (sprout branches) to yield morphologies such as the one in panel (d).

For practical applications in catalysis, the uniformity and details of the nanocluster or nanopillar morphology might not be as crucial as the availability of surface sites which have enhanced catalytic activity, here exemplified by the (111) type symmetry faces. Actually, the issue of how large should the "ideal" (111) surface-portion be for optimal catalytic activity of sites at its center is not fully settled[24,25] and should strongly depend on the specific reaction being catalyzed and on the properties of the substrate. Here we consider the availability of the approximately (111)-coordinated sites within a minimalist definition: All surface sites which are a shared vertex of two equilateral triangles with sides which are nearest-neighbor distanced and which are both in the same plane, were counted as approximately (111)-coordinated. Empirically, we found that just labeling single nearest-neighbor triangles picks too many spurious isolated surface pieces, but the selected two-coplanar-triangle test has yielded a reasonable practical identification method by "covering" those surface regions which were largely (111) type.

For the isolated nanocluster growth at short times, in the plateau regime (defined in Figure 4), the pyramid shaped nanoclusters (e.g., Figure 3) have their side faces largely (111)-coordinated. However, as illustrated in Figure 5, the larger nanopillars, grown at later times, have the (111)-type faces only around the tops (and also near their bases: not shown in Figure 5, but discernable in panel (e) of Figure 1). The vertical sides of the nanopillars predominantly display (100) and (110) type faces. Figure 6 illustrates the time dependence of the areal density, $\theta_{111}(t)$, of the approximately (111)-coordinated surface atoms, the total count of which is $C_{111}(t)$,

$$\theta_{111}(t) = C_{111}(t) / XY. \tag{4}$$

The maximal (111)-type coverage is thus attained for growth times which correspond to the plateau regime of independently grown pyramid-shaped nanoclusters right before they "run into each other." However, in practical situations it may be beneficial to carry out the growth process somewhat beyond the "plateau" times, because once the supply of matter is stopped but before the formed structure is otherwise stabilized, the residual surface diffusion might somewhat erode the formed morphology. Figure 7 shows some features of such growth. Specifically, competing nanopillars grown somewhat past the "plateau" time, while differing in height, will have similar (111) regions near their tops.

In summary, we established that numerical-simulation modeling can yield interesting information on the emergence of the morphology features of growing nanostructures for applications of interest in catalysis. Specifically, growth on flat substrates is best carried out only as long as the resulting structures are isolated nanoclusters. Larger surface features, even if formed as well-defined nanopillars, are not guaranteed to have the desirable surface-face properties. We also found that well-defined surface features are obtained for relatively narrow ranges of parameter values, for those quantities which, here loosely, are related to the on-surface diffusion rate and to temperature. Future studies might explore approaches such as seeding/templating, in order to grow larger (taller) surface features potentially offering larger exposed area with preferred symmetry, such as (111). Furthermore, with substantially more powerful numerical resources invested, it should be possible to carry out realistic simulations with the actual material parameters, as well as for new materials of interest in applications.

We wish to thank Dr. I. Sevonkaev for useful discussions, and acknowledge funding by the US ARO under grant W911NF-05-1-0339.

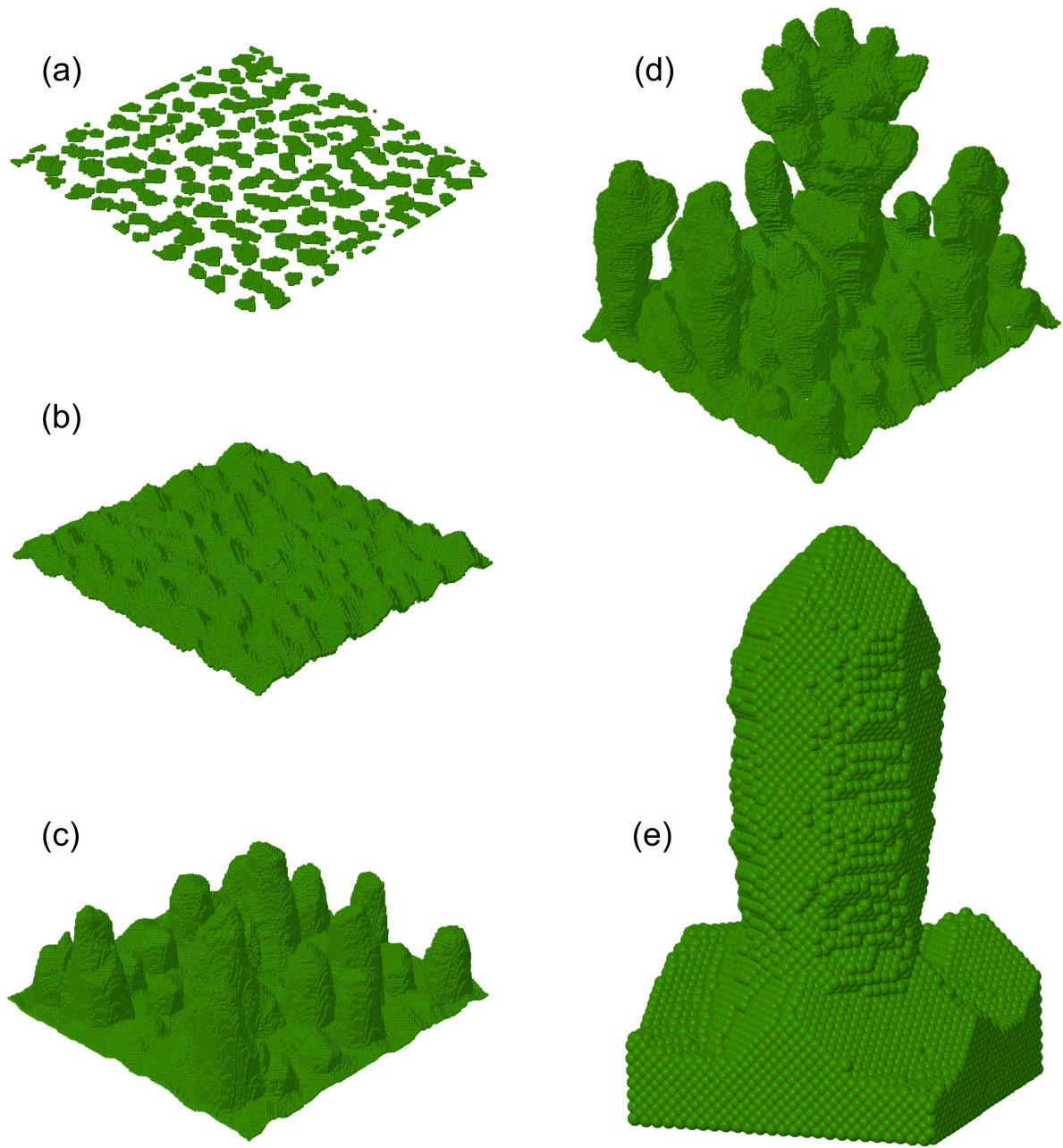

**Figure 1.** Illustration of the emergence of surface-structure nanocluster and nanopillar morphologies obtained in the nonequilibrium growth regime for proper ranges of parameter values, for growth of FCC-symmetry deposits on (100) substrates. Panels (a)-(d): Shown are $200 \times 200$ sections of simulation results actually obtained for $X \times Y = 500 \times 500$, all grown on the initially flat (100) substrates. Only the growing-surface atoms (those that can move or detach) are colored. The parameter values, defined in Section II, here were $n = 1.25 \times 10^{-2}$, $p = 0.6$, $\alpha = 2.5$, and the simulation times were: (a) $t = 0.08 \times 10^6$, corresponding to the initial formation of isolated islands; (b) $t = 0.85 \times 10^6$, corresponding to the stage of emerging pyramidal nanoclusters; (c) $t = 2.44 \times 10^6$, corresponding to the growth of competing nanopillars; (d) $t = 17.14 \times 10^6$, corresponding to the onset of the large-time irregular growth. Panel (e) shows a cutout of a single nanopillar, with the size of the image pedestal $60 \times 60$, with all the non-substrate, deposited atoms colored. This is one of the larger nanopillars obtained for the growth stage shown in panel (c), but taken from another surface portion than that shown in panel (c).

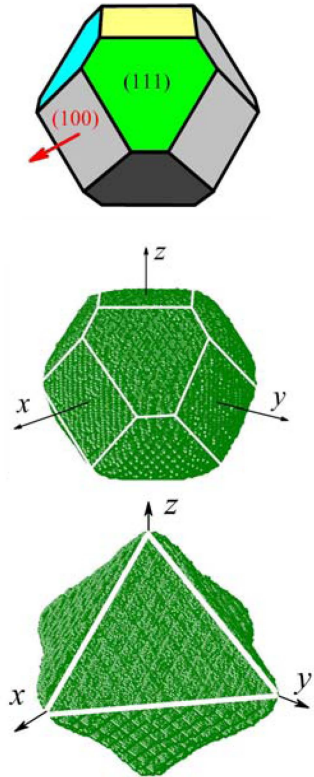

**Figure 2.**  *Top panel:* The Wulff shape for the FCC symmetry, assuming equal interfacial (free-)energy densities of all the faces. *Middle panel:* Nonequilibrium FCC shape obtained under conditions of relatively slow growth.[11] *Bottom panel:* Faster-growth nonequilibrium FCC shape.[11] (The white lines were added for guiding the eye.)

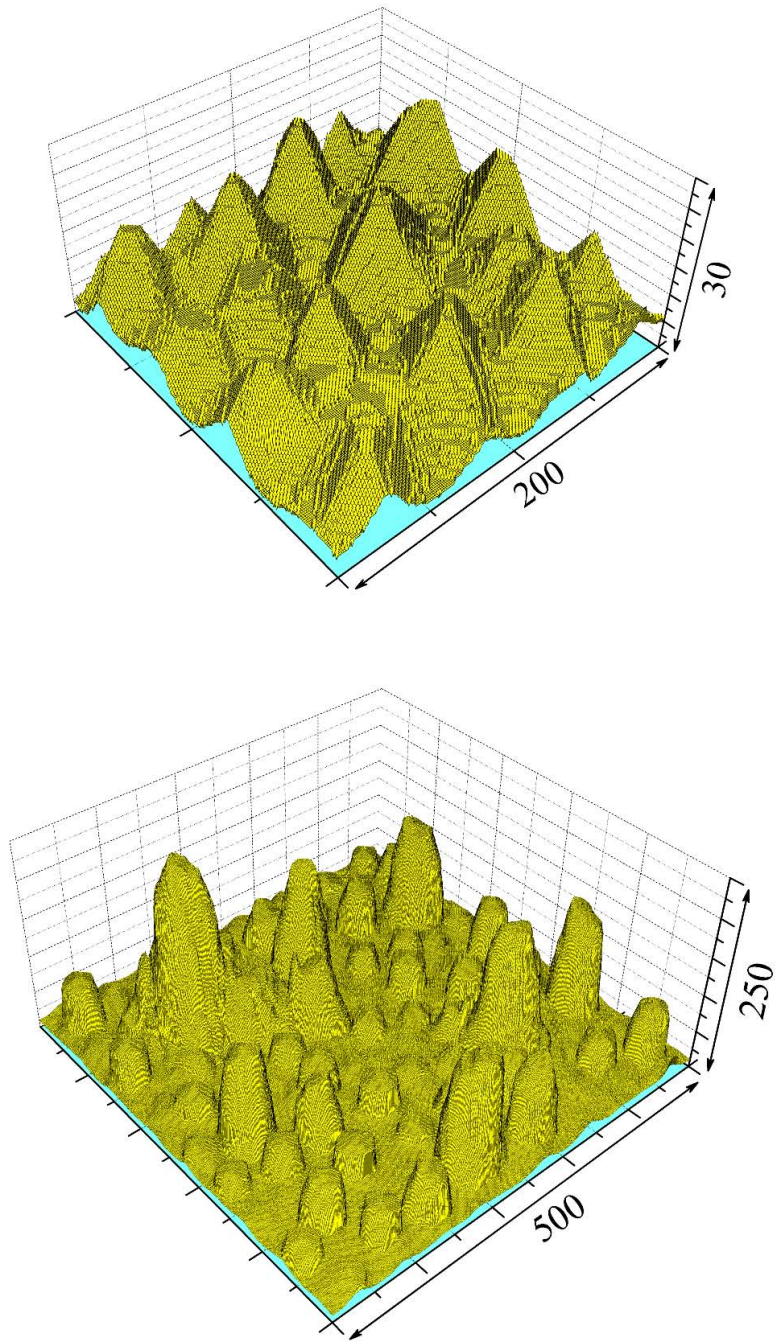

**Figure 3.** *Top panel:* Nanoclusters grown for time $t = 3.5 \times 10^6$, with parameters $n = 8 \times 10^{-3}$, $p = 0.7$, $\alpha = 2.0$. (Shown is a $200 \times 200$ cutout of the simulation results for the $500 \times 500$ substrate, with the vertical scale additionally stretched.) *Bottom panel:* Nanopillars grown by continuing the above simulation to time $t = 30 \times 10^6$.

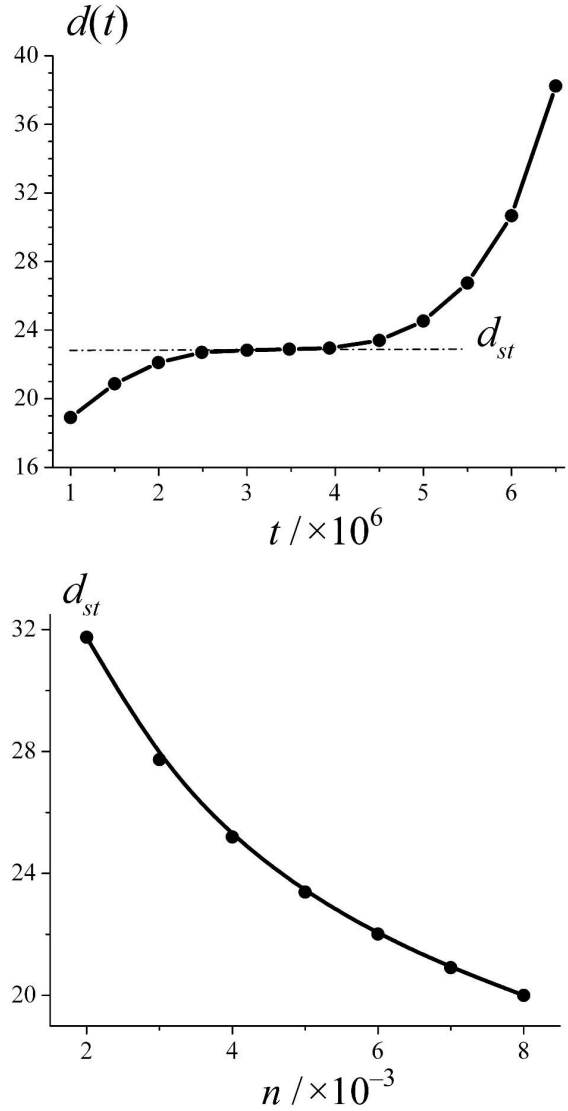

**Figure 4.** *Top panel:* Time dependence of $d(t)$ for the same growth parameters as in Figure 3. (The segments connecting the data points were added for guiding the eye.) The plateau region, with approximately constant $d(t) \approx d_{st}$, is centered at the time corresponding to the morphology of the top panel in Figure 3, and it separates the independent and competitive cluster growth regimes. *Bottom panel:* Variation of the plateau value, $d_{st}$, with the density in the top layer, $n$, which controls the matter flux, with all the other parameters unchanged. The solid line represents the fit to the dependence $d_{st} = \mathrm{const} \times n^{-1/3}$, illustrating that the expected approximate proportionality, $d \sim n^{-1/3}$, indeed holds in the independent-nanocluster growth regime.

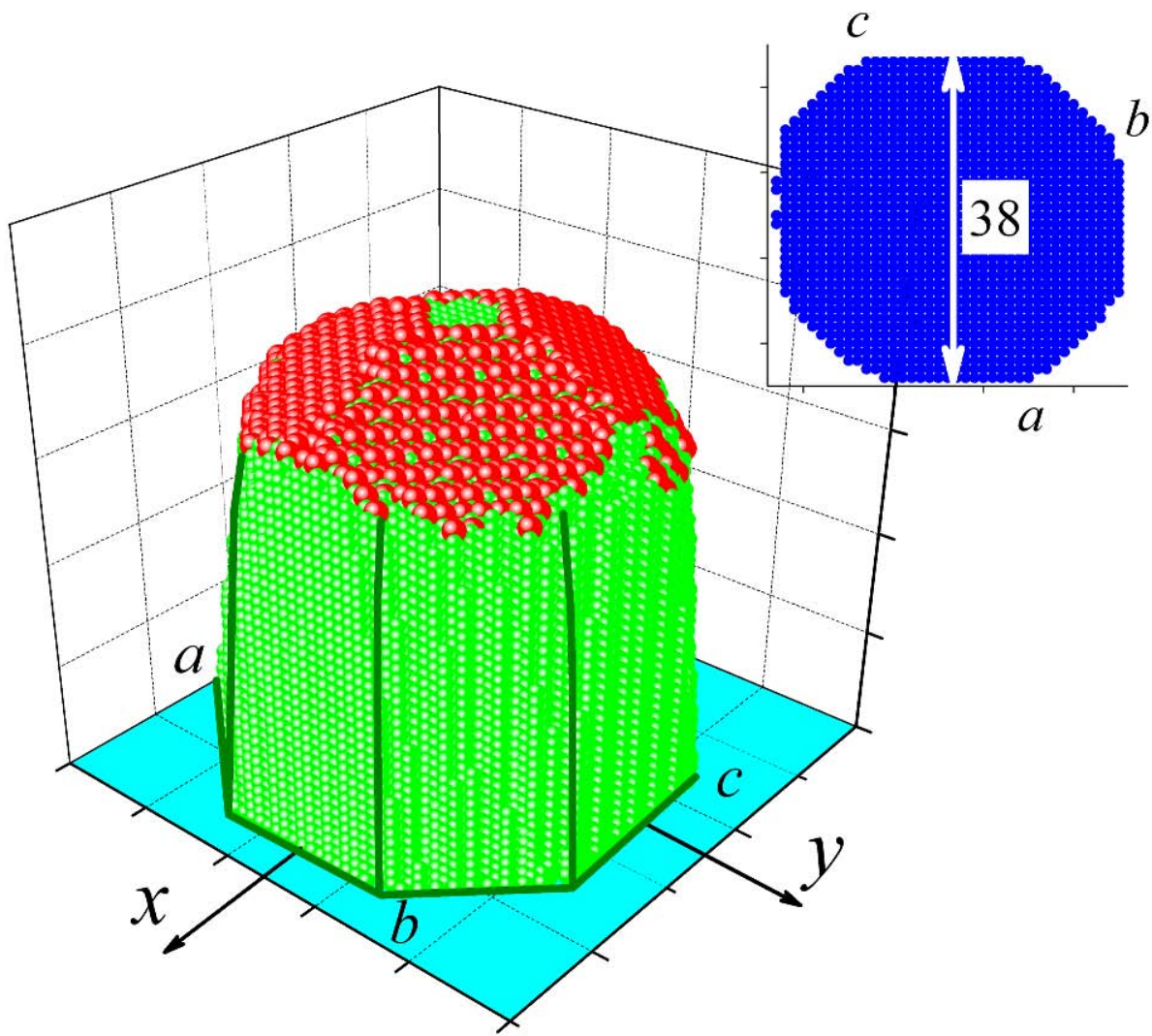

**Figure 5.** A typical single nanopillar reaching height of approximately $z = 62$, cut off its base section: only shown starting from height $z = 20$, grown as part of a deposit formed with parameter values the same as for Figure 3 (for the larger of the two times shown there). The sites identified as approximately (111)-coordinated, as defined in the text, are dressed with red spheres. The other surface atoms are depicted as smaller light-green spheres. The dark-green lines were added for guiding the eye, and a cross-section outlining the octagonal shape of the pillar is also shown as the inset (with the labels *a*, *b*, *c* identifying its orientation). The top "bold spot" is typical for most pillars in the competitive-growth regime and it emerges similarly to the eight corner-truncations by approximately (100)-type faces in the middle panel in Figure 2.

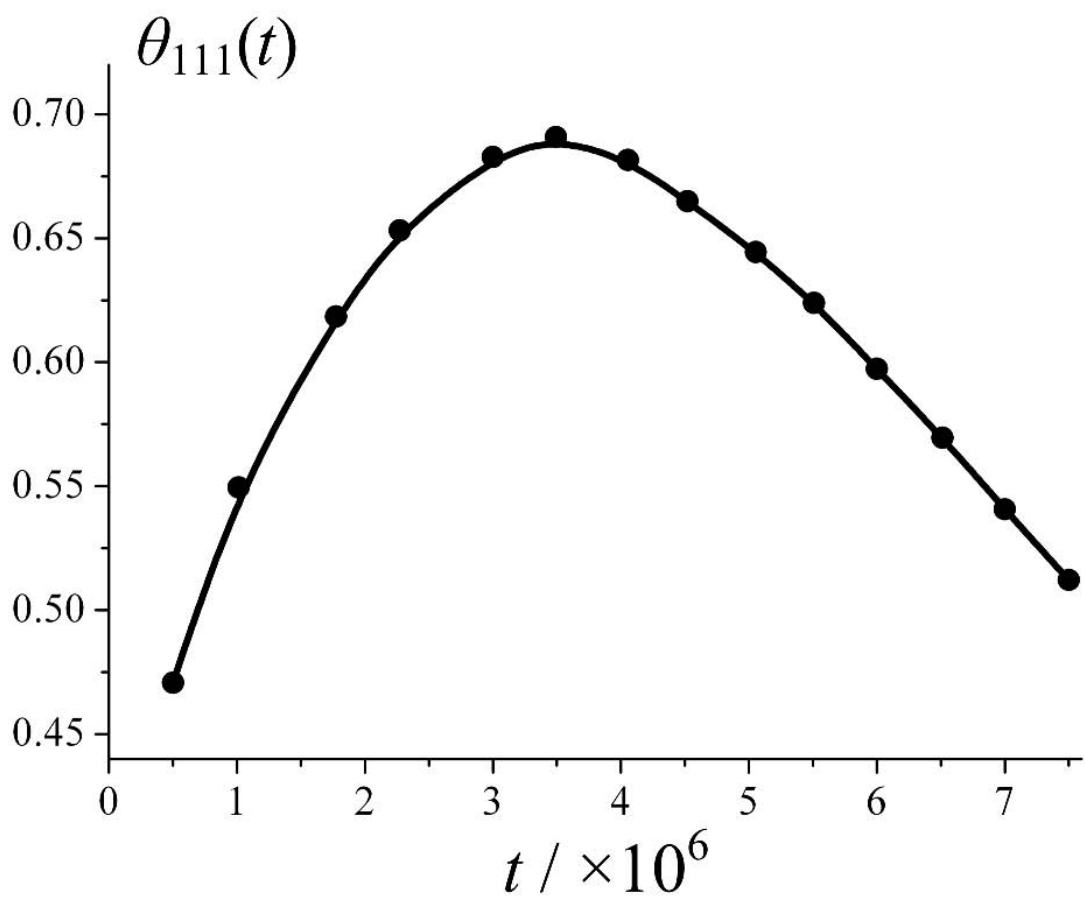

**Figure 6.** Time dependence of the areal density of the approximately (111)-coordinated surface atoms, for the same growth parameters as in Figure 3. (The solid line was added for guiding the eye.)

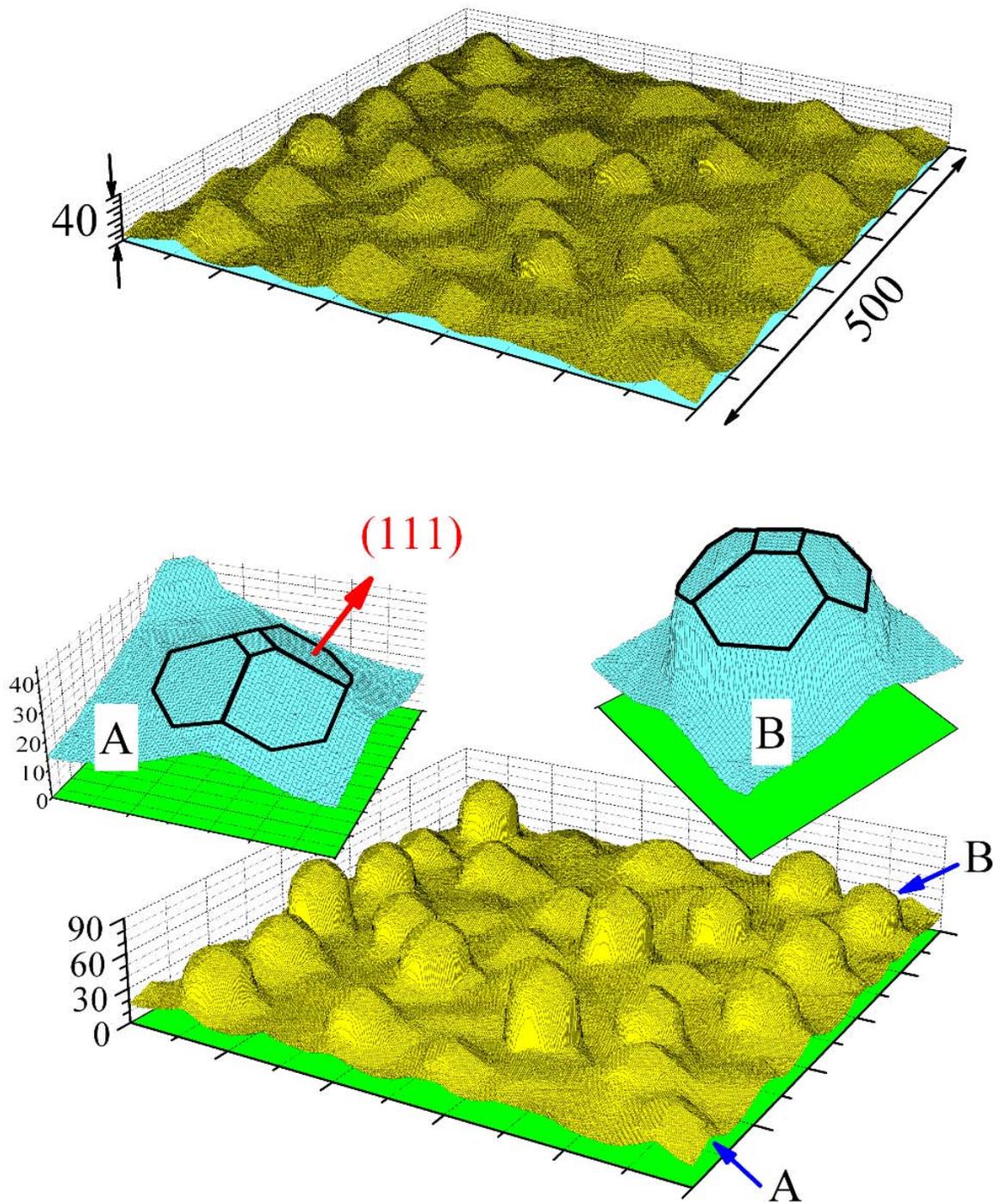

**Figure 7.** *Top panel:* The isolated-pyramid growth stage, for $t = 22.5 \times 10^6$, for parameter values $n = 2 \times 10^{-3}$, $p = 0.7$, $\alpha = 2.0$. *Bottom panel:* Illustration of growth carried out somewhat beyond the "plateau" (isolated-growth) regime, for $t = 27.5 \times 10^6$. The inset diagrams highlight the locations of the hexagonal-shaped, predominantly (111) type regions near the tops of two typical peaks, marked A and B, that grew to different heights at this later growth stage.